\documentclass[12pt,twoside]{article}
\usepackage{fleqn,espcrc1}

\usepackage{graphicx}
\usepackage[figuresright]{rotating}

\newcommand{\AmS}{{\protect\the\textfont2
  A\kern-.1667em\lower.5ex\hbox{M}\kern-.125emS}}
\def\lsim{\raise0.3ex\hbox{$<$\kern-0.75em\raise-1.1ex\hbox{$\sim$}}}
\def\gsim{\raise0.3ex\hbox{$>$\kern-0.75em\raise-1.1ex\hbox{$\sim$}}}
\newcommand{\tr}{{\rm Tr}}
\title{Lattice calculation of medium effects at short
and long distances}

\author{P. Petreczky with O. Kaczmarek, F. Karsch, E. Laermann, 
S.Stickan,
I. Wetzorke \\
and F. Zantow\\
Fakult\"at f\"ur Physik, Universit\"at Bielefeld\\
P.0. Box 100131 33501 Bielefeld}

\begin{document}

\maketitle

\begin{abstract}
We investigate medium effects in QCD like chromoelectric
screening and quasi-particle mass generation by calculating
the heavy quark potential as well as the temporal quark
and gluon Coulomb gauge propagators in quenched approximation. 
\end{abstract}

\section{INTRODUCTION}
\vspace*{-0.2cm}
The perturbative description of the Quark-Gluon Plasma (QGP) 
breaks down even at very high temperatures \cite{Gross81}.
The reason for this is the generation of different length (time)
scales at finite temperature. At high temperatures ($T \gg \Lambda_{QCD}$)
the gauge coupling constant is small $g \ll 1$ and well separated
length scales exist: $1/T \ll 1/(gT) \ll 1/(g^2 T)$.
For such high temperatures the region of applicability of the 
perturbation theory is established.
While for distances smaller than $1/T$ the ordinary perturbation theory
should be applicable, at distances smaller than $1/gT$
the hard thermal loop resummed perturbation theory
\cite{Braaten90} should be applied. 
For distances larger than $1/g^2 T$, however,
perturbation theory is no longer valid \cite{Gross81}.
For interesting temperatures up to a few times $T_c$, however,
$g \gsim 1$ and it is not clear
up to which distances (times) the perturbation theory 
is applicable. At short distances $r \ll T^{-1}$ one expects that
medium effects are not important. For such small distances 
one should also expect the ordinary perturbation theory to 
be viable, because at zero temperature perturbation theory
works at short distances.
Naturally the question arises: at what distances the
medium effects become important and up to which distance 
perturbation theory is applicable? 
In the present work we try to answer these questions by
studying the heavy quark potential and temporal
quark and gluon propagators in Coulomb gauge.

\section{THE HEAVY QUARK POTENTIAL}
\vspace*{-0.2cm}
We consider the color averaged heavy quark potential 
defined by 
\vspace*{-0.15cm}
\begin{equation}
V(R,T)=-T \ln\biggl[{< L(\vec{R}) L^{\dagger}(0)> \over {|<L>|^2}}\biggr].
\end{equation}
\vspace*{-0.15cm}
Here $L(\vec{R})=\tr \prod_{x_0=0}^{N_{\tau}-1} U_0(\vec{R},x_0)$
is the Polyakov loop ($N_{\tau}$ is the temporal
extent of the lattice). So far most studies have concentrated 
on the long 
distance behavior of the heavy quark potential ($RT>1$).
For the physics of heavy quarkonia at finite temperature, however,
it is important to know the potential for $RT<1$. 
In order to explore the short distance behavior of the
heavy quark potential we performed simulations on 
$64^3\times 16$ lattices with the standard Wilson action
at two different temperatures $T=1.5 T_c$ and $T=3 T_c$, as well
as on $32^3 \times 8$ lattices with a tree level Symanzik improved action 
for $T/T_c=1.05,~1.14,~1.23,~1.50,~3.0,~6.0,~12.0$.
The color averaged potential\footnote{
We stick here to the commonly used notion of a potential,
although for $T>0$ one actually calculates the heavy quark free
energy.}
is related to the singlet ($V_1$)and
octet ($V_8$) heavy quark potentials by the well-known 
formula \cite{mclerran81}
\begin{equation}
V(R,T)=-T \ln({1\over 9} \exp(-{V_1(R)\over T})+
{8\over 9}\exp(-{V_8(R)\over T})).
\label{decomp}
\end{equation} 
At very short distances $RT \ll 1$ one would expect that
the singlet and octet potentials are given by their
perturbative zero temperature expressions $V_i=c_i \alpha/R$,
with $c_1=-4/3$ for the singlet and $c_8=+1/6$ for the octet
case. Thus for small distances one would
expect $V(R,T) \sim 1/R$. Expanding Eq. (\ref{decomp}) 
up to $\alpha^2$ one
recovers the well known perturbative result for the color
averaged potential $V(R,T)/T=-\alpha^2/{(3 R T)}^2$
\cite{mclerran81}.
We note, however, that the expansion of  Eq. (\ref{decomp})
couples the perturbative expansion to the high temperature
expansion and therefore the above formula is obviously
not valid at very short distances where $|V_i/T| \gg 1$.
Motivated by the leading order perturbative result
we define an effective coupling $\alpha_{eff}^2(R,T)=-9 V(R,T) T [1/R]^{-2}$.
Here $[1/R]$ denotes the lattice Coulomb potential.
Our results on $\alpha_{eff}(R,T)$ are summarized in Fig. 1a.
One can see that there is a narrow region in $RT$ where
$\alpha_{eff}$ is approximately constant as one would expect
from the leading order perturbative result. We also note
that in this region of $RT$ $\alpha_{eff}(R,T)$ decreases with temperature
as one would expect for the running coupling constant.
For temperatures
$T \le 1.5 T_c$ the effective coupling seems to decrease
at small distances $RT \lsim 0.2$ which may imply that
in this region the potential behaves like $1/R$ according
to the arguments given above. 
For higher temperatures the $1/R$ behavior sets in at smaller
distances because the coupling $\alpha$ is smaller and 
the high temperature expansion works well even at shorter
distances. This probably is the reason that this behavior
so far is not seen in our data for $T \ge 3 T_c$ which are
restricted by the lattice spacing $a=0.0625 T^{-1}$ for
$N_{\tau}=16$ and $0.125T^{-1}$ for $N_{\tau}=8$.
For distances $RT > 0.3$ the
effective coupling decreases with increasing $RT$ which implies the onset
of screening. Thus we can distinguish three different regions
of $RT$ which are  characterized by qualitatively different behavior
of $V(R,T)$: the "true" short distance region where $V \sim 1/R$,
the intermediate perturbative region $V=-\alpha^2/(3 R T)^2$ and
the "large" distance region where screening sets in.
\begin{figure}
\begin{center}
a \hspace*{7.4cm} b\\
\includegraphics[width=18pc]{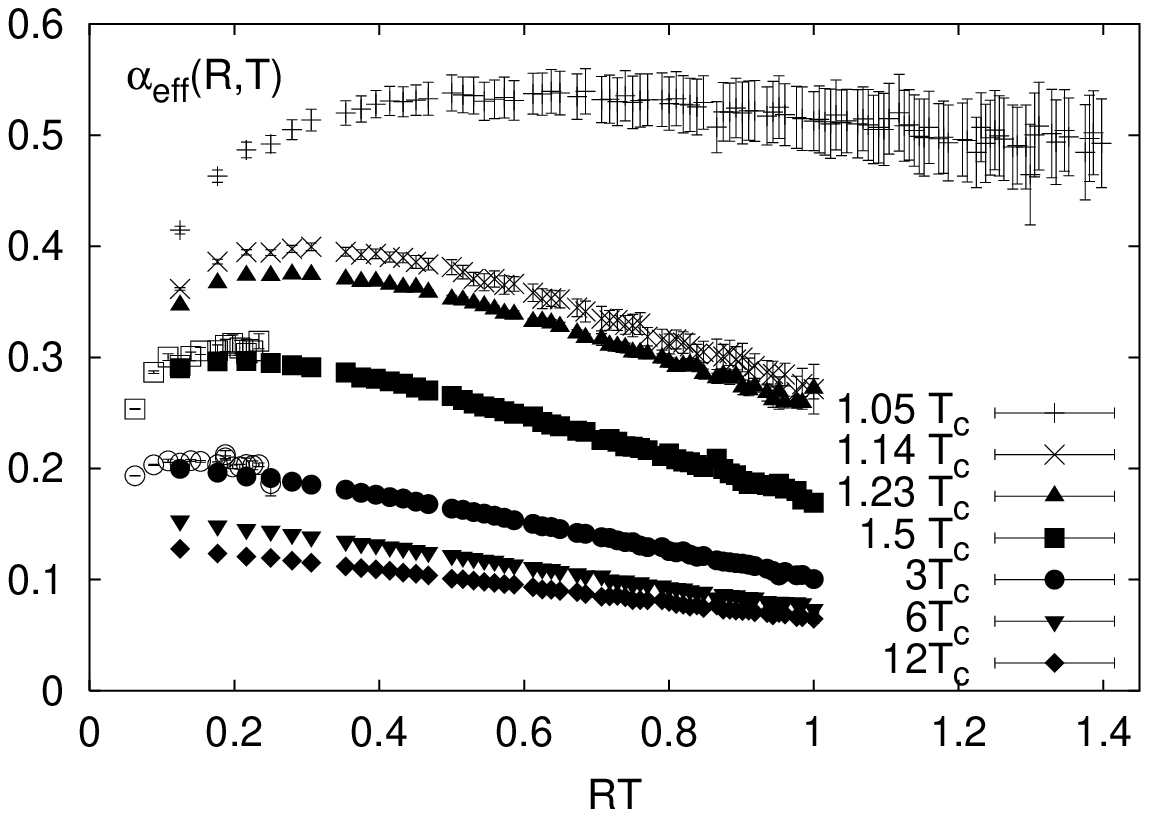} 
\includegraphics[width=18pc]{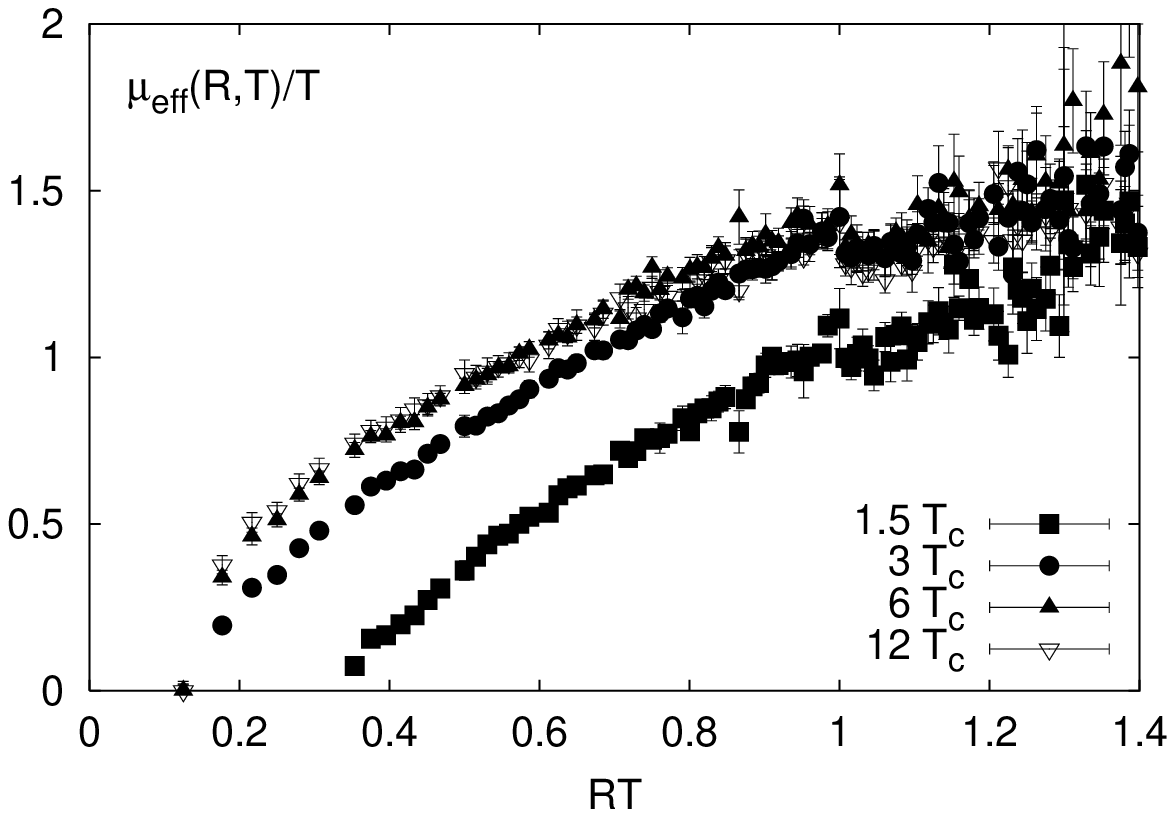}
\end{center}
\vspace*{-1.9cm}
\caption{The effective coupling and the effective
screening masses at different temperatures.
The open symbols in the left figure are the results
from simulations on a $64^3 \times 16$ lattice with the Wilson
action, others correspond to the data obtained on a $32^3\times 8$
lattice with an improved action.
}
\vspace*{-0.9cm}
\end{figure}
We note that for the lowest temperature $T=1.05 T_c$ 
the behavior of $V(R,T)$ is compatible with $1/R^2$ for $RT>0.3$
which may imply that the effect of screening is very small.
To quantify the effects of screening at higher temperatures
it is useful to define an effective screening mass
$\mu_{eff}(R,T)=-\ln(-9 V(R,T) [1/R]^{-2}/\alpha^2(T))/R$ with
$\alpha(T)=\alpha_{eff}(R=0.125T^{-1},T)$.
The behavior of $\mu_{eff}(R,T)$ as function of $R$ is
shown in Fig. 1b. As one can see from the figure
the effective screening masses increase with increasing
distance but up to distances $RT=1.5$ 
do not reach a plateau which can be identified
with the true screening mass.
In fact this behavior of the effective screening masses
could be explained qualitatively in perturbation theory if one
takes into account the momentum dependence of the 
gluon self energy $\Pi_{00}(k)$ \cite{Gale87}.  
Note that the local screening masses in Fig. 1b
are always smaller than screening masses extracted from
the large distance behavior of the heavy quark potential 
in \cite{kaczmarek00} $\mu \sim 3 T$. They are also
smaller than the perturbative Debye screening mass $g(T) T$
if $g$ is defined by $g(T)=\sqrt{4 \pi \alpha_{eff}(R=0.125 T^{-1},T)}$.
We thus conclude that the color averaged heavy quark
potential has a complex structure for interesting distances,
$RT < 1$, and temperatures a few times $T_c$.

\section{THE TEMPORAL QUARK AND GLUON PROPAGATORS}
\vspace*{-0.2cm}
The temporal quark and gluon propagators are related to properties
of quasi-particles in the QGP. 
In fact, in hard thermal loop (HTL) approximation
these propagators are dominated by quasi-particle poles.
\cite{LeBellac96}. The quasi-particle picture of QGP finds
application in resummed perturbative calculations of 
different thermodynamic quantities \cite{Iancu}.
We have performed simulations in quenched QCD with Wilson fermions
on a $64^3 \times 16$ lattice at $T=3 T_c$. 
We have used the standard Wilson action
for the gauge fields and an $O(a)$ improved fermion action (clover action). 
We have calculated the quark and gluon propagator $D_i(\tau,p)$
($i=q,g$) in mixed
$(\tau,p)$ representation ( here $p$ is the absolute value of
the spatial momentum ).

The propagator in this mixed representation can be related to
the spectral function by 
$D_i(\tau,p)=
\int_{-\infty}^{+\infty} d \omega \rho(\omega,p)\exp(-\omega \tau)/(1 \pm 
\exp(-\omega \tau))$. Here ($i=q,g$) and $+(-)$ correspond to the case
of quarks (gluons). From the Monte-Carlo data for $D_i(\tau,p)$
we can extract the spectral functions $\rho(\omega,p)$ with
the help of the Maximal Entropy Method (MEM) \cite{Asakawa00}.
Since the quark and gluon propagator are gauge
dependent quantities one has to fix a gauge. We have fixed
the Coulomb gauge because in this gauge $\rho(\omega,p)$ is
positive which is necessary for MEM. The peaks of the spectral
function $\rho(\omega,p)$ define the dispersion relation $\omega(p)$
which is expected to be gauge invariant. Since the smallest
non-zero momentum $p_{min}=1.57 T$ is relatively large
no detailed information about the properties of collective
excitations like the longitudinal excitation in the gluon sector
or the plasmino excitation in quark sector can be provided
(these excitations exist only in the small momentum region \cite{LeBellac96}).
Here we will not extract the spectral function but postpone that
for future publications. Instead we compare our data for temporal
quark and gluon propagators with the predictions of perturbation
theory in HTL approximation.
In Fig. 2a we show our results for the transverse gluon propagator 
$D_g^T(\tau,p)$ at momenta calculated on 40 gauge fixed configurations. 
We also show there the prediction of the HTL approximation 
using a coupling constant $g \simeq 1.6$ suggested by the short distance
behavior of the heavy-quark potential. 
As one can see from the figure the data disagree substantially with
the HTL prediction even at momenta $p=3.14 T$. We note that
corrections to the HTL approximation will not resolve this
discrepancy since they lead to smaller $\omega(p)$ values shifting
the propagator to values larger than the HTL result 
\cite{schulz94}.

The quark propagator in mixed representation can be
written as $D_q(\tau,p)=\gamma_0 F(\tau,p)+\vec{\gamma} \cdot \vec{n}
G(\tau,p)+H(\tau,p)$ with $\vec{n}=\vec{p}/p$. In the chiral limit
($m_q=0$) $H(\tau,p)=0$. For the value of the hopping parameter $\kappa=0.1339$
(which corresponds to the quark mass close to the chiral limit) 
used in our simulation 
we have found $H(\tau,p)$ to be zero within the statistical accuracy reached in
our simulations. In Fig. 2b we show our results for $F(\tau,p)$
calculated on 40 configurations and 
compared with predictions of the HTL approximation. As in the case 
of the gluon propagator we find large deviations from the
HTL predictions. The situation is similar for $G(\tau,p)$.
We have found that there is no choice of $g$ which can provide
agreement between lattice results and HTL.
We thus conclude that at $T=3 T_c$ medium effects in temporal quark
and gluon propagators are stronger than suggested by HTL perturbation theory.
\vspace*{-0.1cm}
\begin{figure}
\begin{center}
a \hspace*{7.4cm} b\\
\includegraphics[width=18pc]{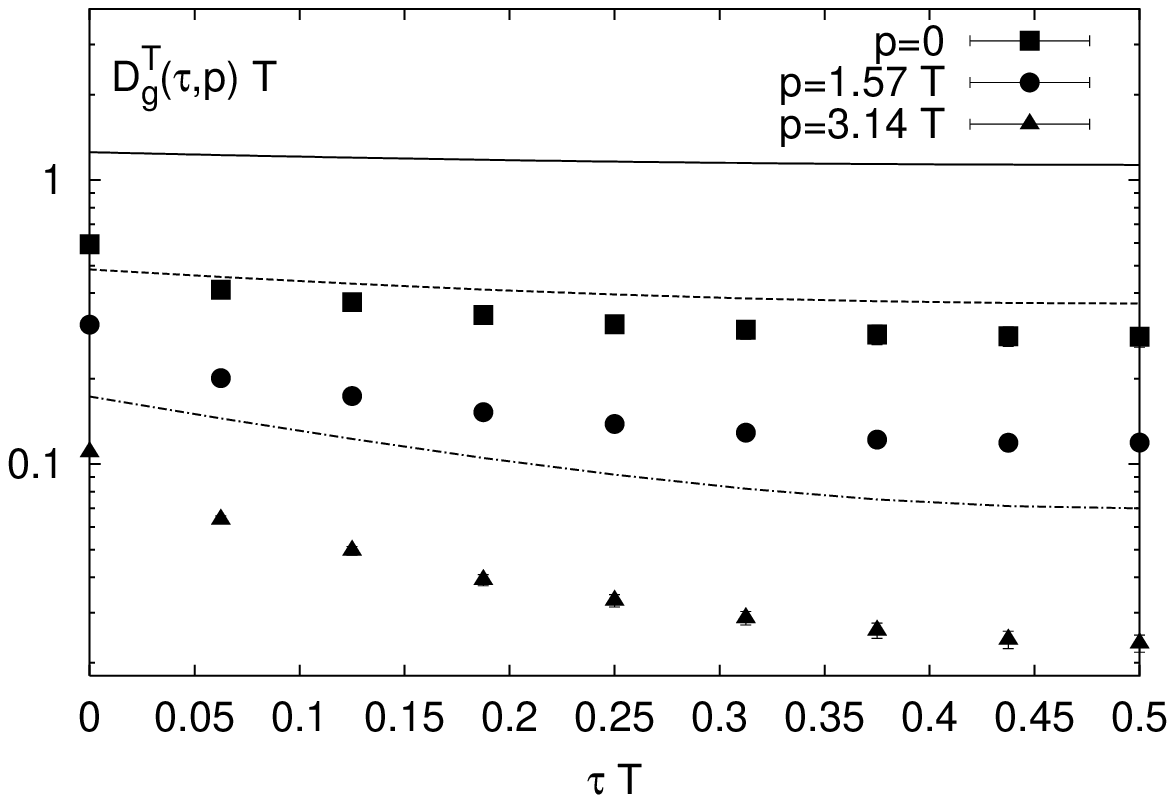} 
\includegraphics[width=18pc]{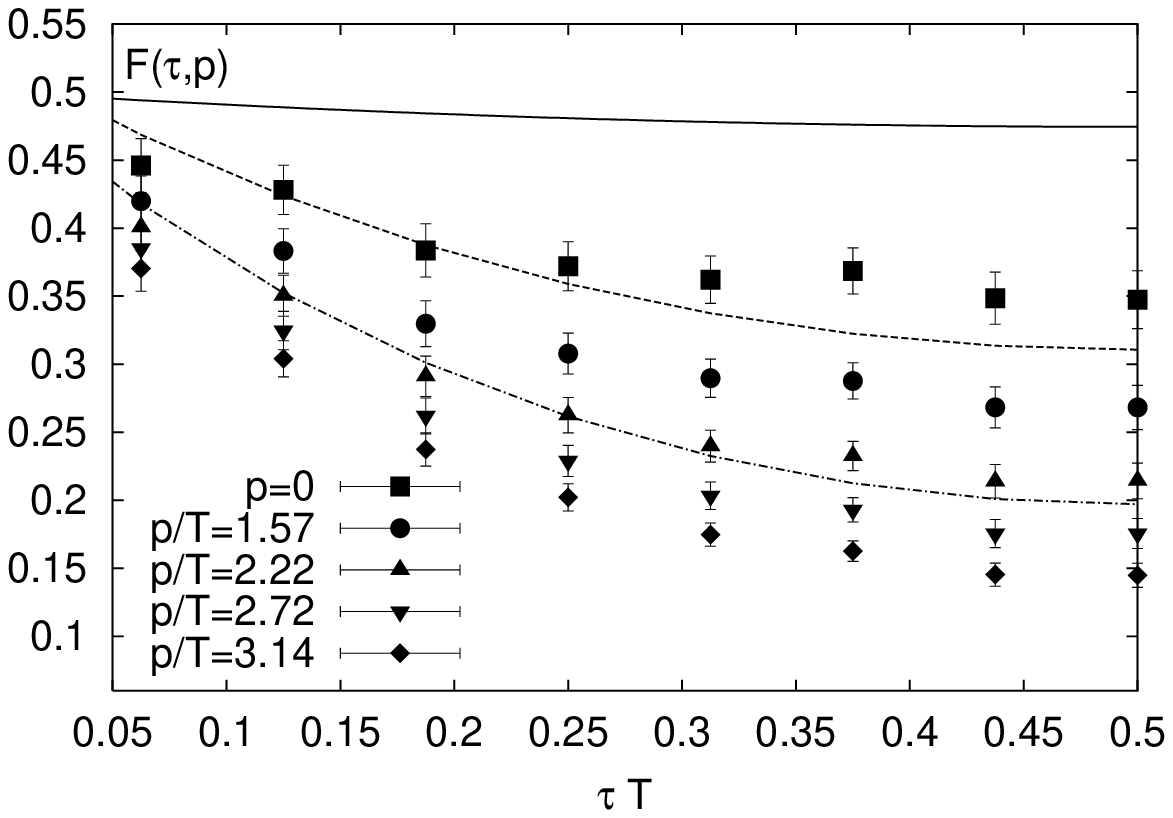}
\end{center}
\vspace*{-1.8cm}
\caption{
The transverse gluon propagator (a) and the
quark propagators (b) for different momenta.
The solid, dashed and dashed-dotted lines 
correspond to the prediction of perturbation theory
in HTL approximation for momenta $p/T=0,~1.57$ and $3.14$
respectively.
}
\vspace*{-0.7cm}
\end{figure}

\end{document}